\documentclass[aps,pra,superscriptaddress,showpacs,twocolumn]{revtex4-1}

\usepackage{amsmath,amssymb,graphicx,color,bm,soul,ulem}
\usepackage[english]{babel}
\usepackage[dvipsnames]{xcolor}
\usepackage{soul}
\usepackage[colorlinks=true,citecolor=blue]{hyperref}
\hypersetup{colorlinks=true,citecolor=blue,linkcolor=blue,urlcolor=blue}

\begin{document}
\title{Non-classical effects in polariton trion
}

\author{T.A. Khudaiberganov}
\affiliation{Department of Physics and Applied Mathematics, Vladimir State University named after A. G. and N. G. Stoletovs, 87 Gorkii st., 600000 Vladimir, Russia}

\author{I.Yu. Chestnov}
\affiliation{ITMO University, St. Petersburg, 197101, Russia}

\author{S.M. Arakelian}
\affiliation{Department of Physics and Applied Mathematics, Vladimir State University named after A. G. and N. G. Stoletovs, 87 Gorkii st., 600000 Vladimir, Russia}

\date{\today}

\begin{abstract}
We investigate quantum phenomena in a system of three coupled microcavities. The possibility of observing polariton blockade in a dimer and triple micropillar configuration is discussed.
The discovered quantum effects allow using these systems as versatile sources of individual polariton photons. Various manifestations of the quantum blockade can be tuned with the use of the pumping laser frequency. 
We discovered that the action of an artificial gauge field on a polariton trion causes the effect of a collective quantum blockade
-- a phenomenon consisting in blocking of excitation of the state with $n$ particles distributed over multiple coupled modes.
We found that when a collective quantum blockade on a non-Hermitain polariton dimer as part of the trion and a blockade on the machine itself with an antibunching effect of a micropillar coupled to the dimer, then a polariton dimer is entangled with that micropillar.

\end{abstract}

\maketitle

\section{introduction}

The quantum nature of light is reflected in its statistical properties, such as sub-Poissonian statistics, and in the violation of classical inequalities for correlation functions. One such phenomenon is photon blockade, where the interaction of a single photon with a nonlinear system prevents interactions with additional photons, resulting in the emission of only one photon. This effect enables systems to suppress the emission of photon pairs, leading to sub-Poissonian statistics of radiation \cite{Klyshko1996}. Photon blockade \cite{Bennett2005, Zubizarreta2020, Sanvitto2019, Rabl2011, Miranowicz2013} is a key mechanism for converting coherent laser light into a stream of individual photons in a Fock state \cite{Flayac2013}, accompanied by the emergence of autocorrelations in photon counts.

The quantum anticorrelations of radiation emitted by nonlinear systems typically arise from the anharmonicity of the spectrum \cite{Bennett2005,Paul1982}. This anharmonicity causes the first excited state of an oscillator to be non-resonant with the second excited state, making the simultaneous appearance of two photons at the same frequency unlikely (although this becomes possible with dissipation-induced broadening). A pronounced photon blockade effect generally requires an interaction strength $U$ comparable to the total loss rate $\gamma$. In contrast, unconventional photon blockade (UPB) can occur in weakly nonlinear systems with $U < \gamma$, a condition typical in condensed matter physics \cite{Zubizarreta2020,Shen2014,Snijders2018,Flayac2017}. To maintain a stationary population of optical modes, resonators are commonly used. Low-dimensional systems, such as quantum wells \cite{Schneider2018} or quantum dots \cite{Ningyuan2018} embedded in microcavities, are typical setups for studying light-matter interaction.

A key feature of nonlinear systems that can exhibit quantum correlation is their potential to generate entangled states. These states, a form of quantum correlation between subsystems, have no classical counterpart and are characterized by the non-separability of quantum states. In this context, entanglement plays a crucial role as a non-local resource in various quantum technologies, including quantum computation, communication, and cryptography. Entangled states are particularly unique in that knowledge of the entire system does not necessarily provide full information about the individual subsystems, further highlighting the complexity and power of quantum correlations.


The entanglement hierarchy of multiparticle mixed states \cite{Cirac2000} indicated which partitions of the composed system into parts can be separability. A multipartite state is called $k$-separable if it can be represented as a $k$-partite state for some partition into $k$ parts. By treating each group as a single entity, a previously entangled state may become completely separable with respect to a coarser partitioning. 

For continues variables the entanglement can be determined using a special Hermitian operator ("witness") $W$. This operator has a negative expected value $Tr(W\rho)=\langle W \rangle <0$, as shown in \cite{Barbara2000}. If such an operator is found, the state is entanglement. 
For continuous observables, Hillery and Zubairi derived inequalities on quantum observables of angular momentum operators
\cite{Hillery2006}.
Similar inequalities are related to the non-classical properties of quasiprobability in the quantum phase space \cite{Miranowicz2010}. 
A collective bosonic state is considered non-classical if the Glauber-Sudarshan $P$-function cannot be interpreted as a classical probability density, i.e., it is non-positive or more singular than the Dirac delta function \cite{Miranowicz2010}. If $\langle:\hat{f}^{\dag{}}\hat{f}:\rangle<0$, where $\hat{f}-$ is a function from the create and annihilation operators, and $::$ denotes normal ordering, then the condition is non-classical. 

This research aims to investigate the relationship between quantum blockade and entanglement in polariton systems. Quantum blockade can be viewed as a mechanism that prevents certain states from being populated, a concept that bears resemblance to quantum entanglement. For example, in a two-qubit system, the entangled state $\frac{|00\rangle + |11\rangle}{\sqrt{2}}$ we can imaging as blockade the states $|10\rangle$ and $|01\rangle$, in the sense that these states cannot appear in the superposition. This analogy suggests that quantum blockade could influence the formation or manipulation of entangled states, and understanding this connection may provide valuable insights into quantum correlations and enhance applications in quantum technologies.

Semiconductor Bragg resonators are capable of focusing optical radiation into a small volume on the micrometer scale \cite{Bork1991}. These structures also feature high $Q$-factors, enabling control over photon signals that is not possible with other types of resonators. Manufactured using molecular beam epitaxy, the most common configuration is flat resonators, which resemble quantum wells sandwiched between two Bragg mirrors \cite{Oesterle1995}. Chemical etching allows precise shaping of these resonators, achieving high optical radiation concentration in the in-plane direction. An example is resonators with a circular cross-section, such as pillar microcavities (MPs) \cite{Brichkin2011,Zhang2014,Reitzenstein2010}, which typically exhibit weak nonlinearity \cite{Ferrier2011} with $U \approx 0.01\gamma$. These micropillar systems enable the generation of highly efficient single-photon sources with $g^{(2)}$ values as low as 0.05 \cite{Gazzano2013,Somaschi2016}. The sub-Poissonian statistics of radiation emitted from MPs, including those supporting strong coupling between exciton and photon modes, has been theoretically predicted for both weak \cite{Verger2006,Eleuch2008} and strong pumping \cite{Demirchyan2017,Khudaiberganov2022}.

\begin{figure}
\includegraphics[width=0.49\textwidth]{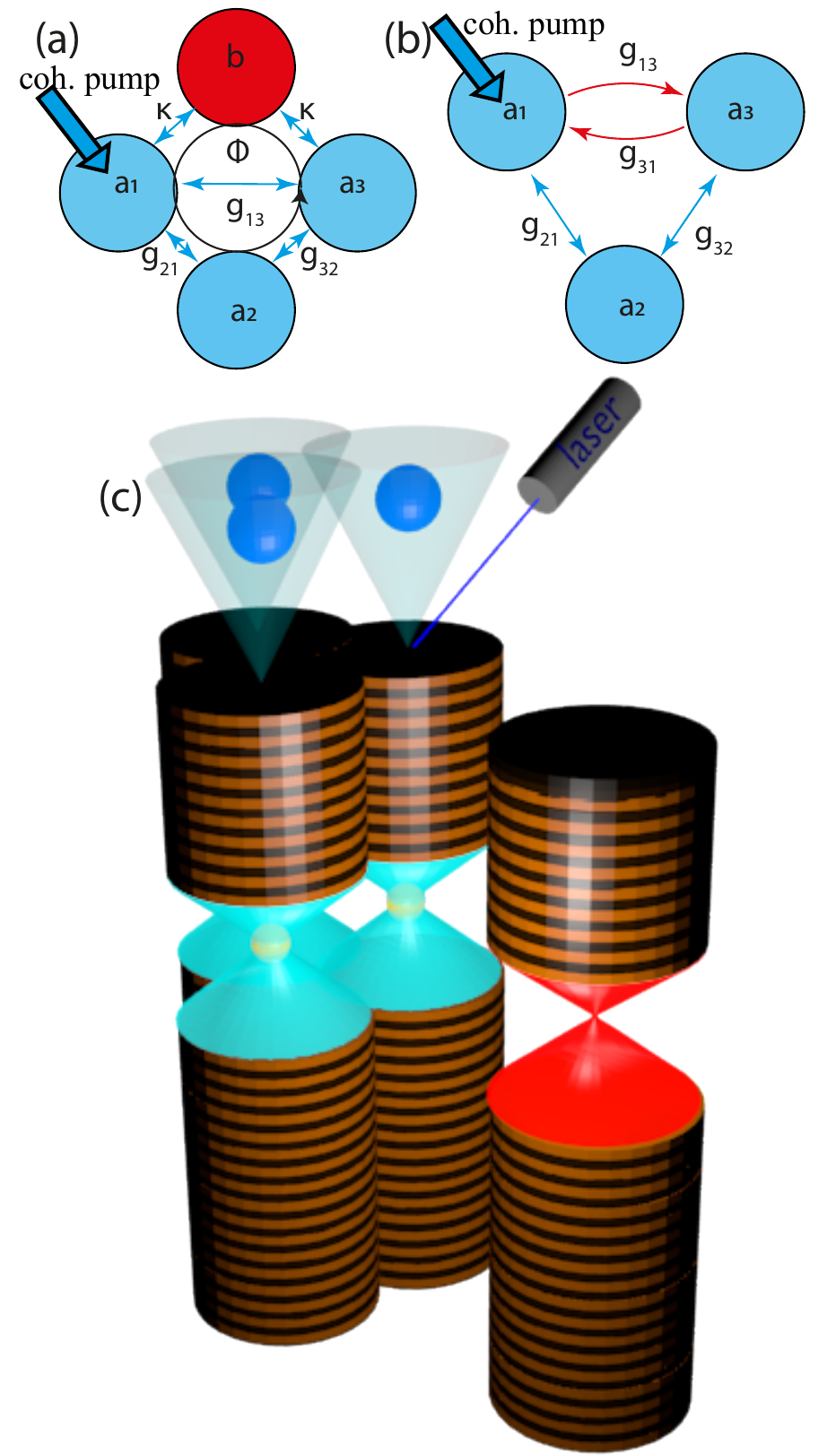}
\caption{ (a) Schematic of polariton quadruplet;
(b) The scheme of polariton trion as an model in which the rapidly decaying mode «b» of the fourth micropillar (empty MP, i.e., without an active medium) are coupled with the first and third amplitude $\kappa$ is adiabatic elimination;
(c) The scheme of design polariton quartet - “trion+empty MP”.}
\label{fig1}
\end{figure}

In recent years, non-Hermitian physics with conserved energy have attracted a lot of attention \cite{Song2021,Huang2022,Lei2024}.
In particular, in
$PT$-symmetry with systems the effects of quantum blockade have already been found at use of nonreciprocal coupling \cite{Huang2022}.
In work \cite{Clerk2022} obtained nonreciprocal coupled between two quantum system use the highly damped mode in our notate denote as «b». In this article, we add another micropillar to the non-Hermitian dimer and introduce another “artificial” gauge field see Fig.~\ref{fig1}(a) which can be thought of as a nonreciprocal coupled Hermitian trion see Fig.~\ref{fig1}(b). Model of the polariton quartet are shown in Fig.~\ref{fig1}(c).
For a quartet of micropillars having two cycles \cite{Lannes2011}, see Appendix A. 
These phases can be obtained by surface acoustic waves modulating the eigenfrequency of the micropillar \cite{Villa2017}.

The article has the following structure: in the first section, we obtained a general model of a polariton trion consisted of three closely placed MPs. In the second section, we consider the statistics of a polariton dimer, further we considered the statistics of a polariton trion in the calibration-invariant phase. Third section obtain condition for collective quantum blockade of a polariton trion and considered entanglement effects.

\section{Model}

The physical basis for the study of multiparticle interactions can be represented by array of micropillars interacting with each other via tunneling photons. 
In this paper, we consider a quartet of micropillars (three micropillars with polariton modes and low-quality micropillar without an active medium), which is reduced to a trion consisting of a non-Hermitian dimer «1+3» and micropillar «2» see Fig.~\ref{fig1}(b). We investigate the non-trivial statistical properties of the radiation of such a system in the presence of an artificial gauge field.

\begin{figure}
\center{\includegraphics[width=\columnwidth]{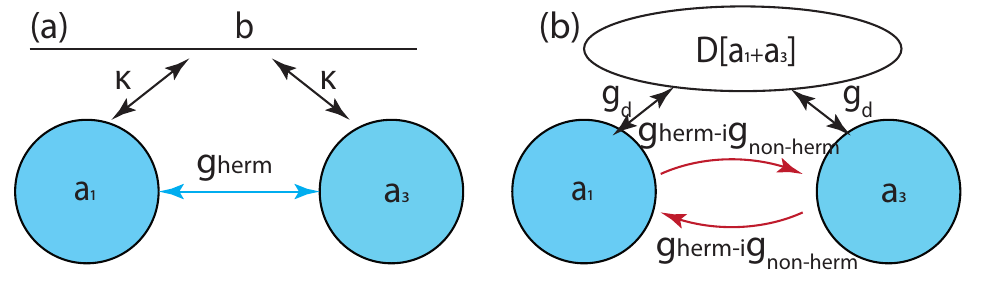}}
\caption{(a) The diagram illustrates the connection polaritons into a dimer (13) with the fast damped photon mode «b» (and with coupling amplitude $\kappa$); (b) Approximate model of scheme (a) as a non-Hermitian dimer with nonreciprocal complex coupling and non-local dissipator \cite{Clerk2022}.}
\label{fig2}
\end{figure}

Let us consider the following polariton quartet model:

\begin{equation} \label{Eqm2_}
\begin{array}{l} {\hat{H}=\mathbf{\hat{a}^{\dag}} \mathbf{G^{(1)}}\mathbf{\hat{a}}+\sum_{i=1}^{3}{F_{i}(\hat{a}_{i}^{\dag}+\hat{a}_{i})+U\hat{a}^{\dag}_{i}\hat{a}^{\dag}_{i}\hat{a}_{i}\hat{a}_{i}}}, \end{array}
\end{equation}
here $\mathbf{\hat{a}} =\left( \hat{a}_1,\hat{a}_2, \hat{a}_3,\hat{b}\right)^\intercal$, $\mathbf{\hat{a}^{\dag}} =\left( \hat{a}_1^{\dag},\hat{a}_2^{\dag}, \hat{a}_3^{\dag},\hat{b}^{\dag}\right)$ the vector of annihilation and creation operators of polaritons (bosons) of the first, second, third micropillars and the fourth low-quality and “empty” (without active medium) micropillar, which we will denote as «b»; $U-$ the cubic nonlinearity parameter (due to elastic exciton-exciton scattering); $F_i-$ is the amplitude of coherent pumping of the $i^{\rm th}$ micropillar; $G-$ is the matrix of the quadratic form of the linear part of the Hamiltonian, which has the following form:
\begin{equation} \label{Eqm3_}
    \mathbf{G^{(1)}}=\begin{bmatrix}
    \Delta_1& g_{12} & g_{13} & \kappa\\
    g_{21}& \Delta_2 & g_{23}& 0\\
    g_{31}& g_{32} & \Delta_3 & \kappa\\
    \kappa& 0 & \kappa & \Delta_b
\end{bmatrix},
\end{equation}
here $\Delta_i=\omega_i-\omega_L$ detuning $i^{\rm th}$ of micropillar from laser frequency $\omega_L$ with amplitude $F_i$ for a given $i^{\rm th}$-micropillar,  $g_{ij}$ the amplitude (tunneling) coupling between $i^{\rm th}$ and  $j^{\rm th}$ micropillars in polariton «molecule» by photons tunneling through the walls of the micropillar \cite{Kaliteevskii1997}. 
The calibration-invariant phase is eliminated by the gauge transformation $a \rightarrow a e^{i\phi}$ only if the system is not closed in a loop.
The micropillar connection diagram is shown in Fig.~\ref{fig1}(a) has two cycles, there are two invariant phases $\Phi_1,\Phi_2$ \cite{Lannes2011}, see Appendix A.
Let take two coupled modes with complex frequencies $\omega_1-i\gamma_1/2$ and $\omega_3-i\gamma_3/2$, which are connected by an Hermite coupling $g_3$ and each of which is coupled with a rapidly damped mode $b$ with coupling coefficient $\kappa$ with dissipation parameter $\gamma_b$, and $\gamma_b\gg 
\gamma_{1,3}$, see Fig.~\ref{fig2}.

In adiabatical approximation excluding the highly damped mode «b» we obtain the non-reciprocal coupled of oscillators «1+3» and non-local dissipation with dissipator of common bath \cite{Clerk2022}:




\begin{equation} \label{Eqm10_}
D[b] \approx |\kappa|^2 D[a_1+a_3],
\end{equation}
in our notation $g_{d}=|\kappa|^2$, $\gamma_{1,3}=\gamma_{LP1,3}-i\Gamma$, $g_{+}=g_3,g_{-}=\Gamma$ and $\Gamma=|\kappa|^2/\gamma_b$.

Thus, we can exclude mode  «b» from the system and obtain the following form of matrix \eqref{Eqm3_}:


\begin{equation} \label{Eqm11_}
    \mathbf{G^{(1)}}=\begin{bmatrix}
    \Delta_1& g_{12} & e^{i\Phi_1}g_{+}-\frac{i}{2}e^{i\Phi_2}g_{-}\\
    g_{21}& \Delta_2 & g_{23}\\
    e^{-i\Phi_1}g_{+}-\frac{i}{2}e^{-i\Phi_2}g_{-}& g_{32} & \Delta_3
\end{bmatrix},
\end{equation}
here $g_{\pm}$ -- Hermitian and non-Hermitian coupling amplitude. $\phi_{1,2}$ -- gauge-invariant phases.

The matrix \eqref{Eqm11_} describes the linear part of the Hamiltonian of the trion consisting of the non-hermitian dimer «1+3» and the micropillar «2», they are connected in a triangle and are in “artificial” gauge fields (note that there is only one invariant phase in the triangle, but because of the nonreciprocity of the coupling there can be two invariant phases in the system).

To describe the evolution of the open quantum we will use the master equation (Lindblad equation) on the density matrix:

\begin{equation} \label{Eqm12_}
\begin{array}{l}
{i\frac{\partial \rho }{\partial t} =\left[\hat{H}_{t}^{+},\rho\right]+\{\hat{H}_{t}^{-},\rho\}+i\sum_{j=1}^{3}{\gamma_{j} \mathbf{D} [\hat{a}_{j};\rho]}}\\{+ig_{d} \mathbf{D}[\hat{a}_1+\hat{a}_3;\rho]},\end{array}
\end{equation}
here $\{.,.\}$ denotes the anticommutator; $\hat{H}_{t}^{\pm}$ are the Hermitian and non-Hermitian parts of the Hamiltonian, respectively. In \eqref{Eqm12_} we defined the dissipator superoperator $\mathbf{D}[a;\rho]=\hat{a}\rho\hat{a}^{\dag}-\frac{1}{2}\{\Hat{a }^{\dag}\hat{a},\rho\} $ which describes the interaction with the environment with relaxation rates $\gamma_{1,2,3}$ for the first three micropillars.
Last term in \eqref{Eqm12_} is non-local dissipator.

The equation (\ref{Eqm12_}) is the most accurate description of relaxation processes in open quantum systems. However, in the case of weak pumping regime, we can neglect quantum jumps $a_i\rho a_j^{\dag}$ (where $i,j=1,2,3$ and $i \neq j$), we can restrict ourselves to solving the Schr{\"o}dinger equation with the Hamiltonian \eqref{Eqm15_}.

When we have low pump ($F<\gamma$) we can neglecting the annihilation part of the pump Hamiltonian for weak pump limit due to sub-leading ordered $c^{(n+1)} \ll c^{(n)}$ \cite{Wang2024}. Then we write the follow block-matrix Hamiltonian
\begin{equation} \label{Eqm15_}
H\approx diag(1, G^{(1)}, G^{(2)},...,G^{(n)})+diag(f^{(1)}, f^{(2)},...,f^{(n-1)})
\end{equation}
the notation $diag(.)$ is a diagonal block-matrix. 
The first part of the Hamiltonian is a block one and consists of  matrices $G^{(n)}$ acting on $n$-particle states with rank n(1+n)/2, these can be represented as tensor networks, see appendix B. The second part of the Hamiltonian takes the system from a lower rank tensor network to a higher one.

The first term in the Hamiltonian \eqref{Eqm15_} acting on a $n-$particle state use the matrix $G^{(n)}$, distributed $n-$ particles over multiple modes, the second term in \eqref{Eqm15_} reduces the number of particles in the system, changing the system to a tensor state with $(n- 1)-$ particles by the rectangular matrix $f^{(n)}$.


We account for non-Hermiticity in Hamiltonian \eqref{Eqm15_} in Ha by extending the domain of definition of the detuning parameters to complex numbers (meaning that the energy is not conserved but the number of particles is conserved): $\Delta_i \Rightarrow \tilde{\Delta_{1,3}}=\Delta_{1,3}-i \gamma_{1,3}/2 -ig_{-}/2, \tilde{\Delta_2}=\Delta_2-i \gamma_2/2$.




\section{Methods}

The task was to identify the optimal parameters at which the quantum blockade effect would be most pronounced. To achieve this, we first determined the stationary amplitudes as defined in equation \eqref{Eqt5_} when the correlation functions are minimal. We then employed the QuTiP package, version 4.7.1 of the Quantum Toolbox in Python \cite{qutip}, to numerically solve the master equation \eqref{Eqm12_} for the parameters takes from solution of \eqref{Eqt5_}.

\section{Results}

The following consideration is of few coupled quantum anharmonic oscillators in a thermal bath. If energy is pumped into one of the systems (for example, by laser radiation), it is possible to establish equilibrium and set the system in a stationary state.
Let us make the following assumptions: transitions between states for different subsystems are only possible if the total number of particles is conserved; the number of particles in the system will change only because of coherent pumping. From the Schr{\"o}dinger equation, the probability amplitudes of the system being in one or another state can be found.

We start by considering the quantum statistics of the dimer studied in \cite{Flayac2017}. Further, we connect another micropillar with different coupling amplitudes to this dimer. Next we introduce a gauge phase. Then we add non-Hermitianity to the polariton trion system on the original polariton dimer proposed in the work \cite{Clerk2022}.

\subsection{Dimer}

\begin{figure}
\center{\includegraphics[width=\columnwidth]{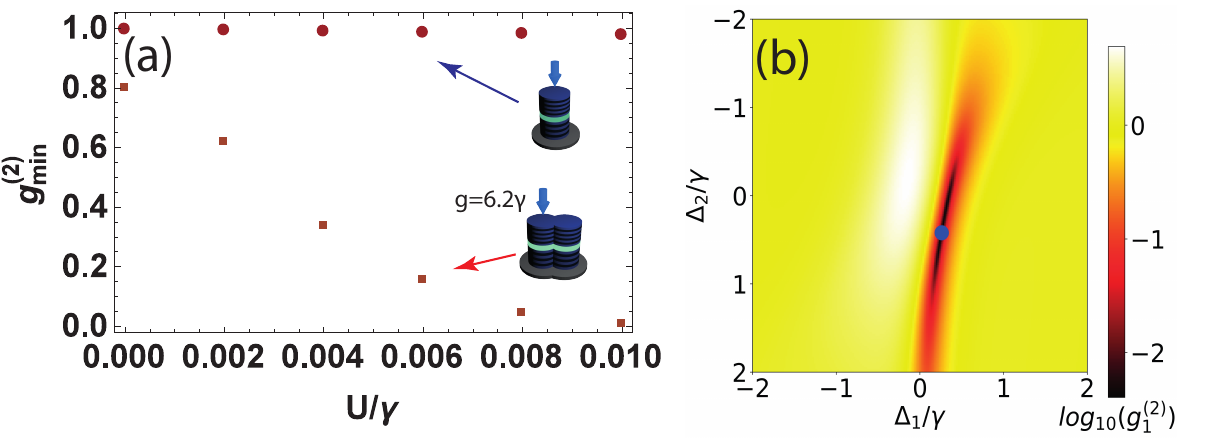}}
\caption{(a)  The minimum value of the second order correlation function from $U/\gamma$ for the first micropillar (which is pumped up) $g^{(2)}_2$ for a single micropillar and for dimer with  coupling parameter that corresponds to the minimum value of the second moment for the value of the nonlinearity parameter $U/\gamma=0.01$ equal to $g_{13}=6.2\gamma$ obtained using \eqref{Eqd11a_}.
Other parameters are: $U_1=U_2=U_3=U_{LP}=0.01\gamma,\gamma_{1}=\gamma_{2}=\gamma_{3}=\gamma,F_1=\gamma_1,F_2=0,F_3=0,\hbar\gamma=0.0274$~meV. (b) The correlation function map for the first micropillar $g^{2}_1$ in dependency of detunings $\Delta_{1}/\gamma$ and $\Delta_{2}/\gamma$ for the optimal coupling constant $g_{13}=6.2\gamma$. The blue disk (with coordinates $\Delta_1/\gamma=0.3,\Delta_2/\gamma=0.1$)
on the map corresponds to the minimum value of the function $g^{(2)}_1$.}
\label{fig3_1}
\end{figure}


A single micropillar exhibits only weak non-classical behavior $g^{(2)}\approx1$ , as illustrated in  Fig.~\ref{fig3_1}a. In the case of dimer - two coupled micropillars see Fig.~\ref{fig3_1}a, we can can be observed the sub-Poisson statistics for small nonlinearities $U/\gamma<1$ based on the unconventional blockade mechanism \cite{Flayac2013}. 
The optimal conditions for the minimum of $g^{(2)}_1 \approx 0.01$ are shown in Fig.~\ref{fig3_1}b for the quantum blockade regime.

The condition 
of the optimal quantum blockade \cite{Flayac2017} (for the resonance case) for the coupling parameter
\begin{equation} \label{Eqd11a_}
g_{opt}/\gamma \approx \sqrt{\frac{2}{27}} \sqrt{\sqrt{3+4\left(\frac{U}{\gamma}\right)^2}\left(\frac{U}{\gamma}+3 \frac{\gamma}{U}\right)-2 \left(\frac{U}{\gamma}\right)^2}
\end{equation}

and detuning

\begin{equation} \label{Eqd11b_}
\Delta_{opt}/\gamma \approx -\frac{1}{3}\frac{U}{\gamma}+\frac{1}{6}\sqrt{4\left(\frac{U}{\gamma}\right)^2+3\left(\frac{\gamma}{U}\right)^2}.
\end{equation}

\subsection{The antibunching effect in Hermitian trion}

\begin{figure}
\center{\includegraphics[width=\columnwidth]{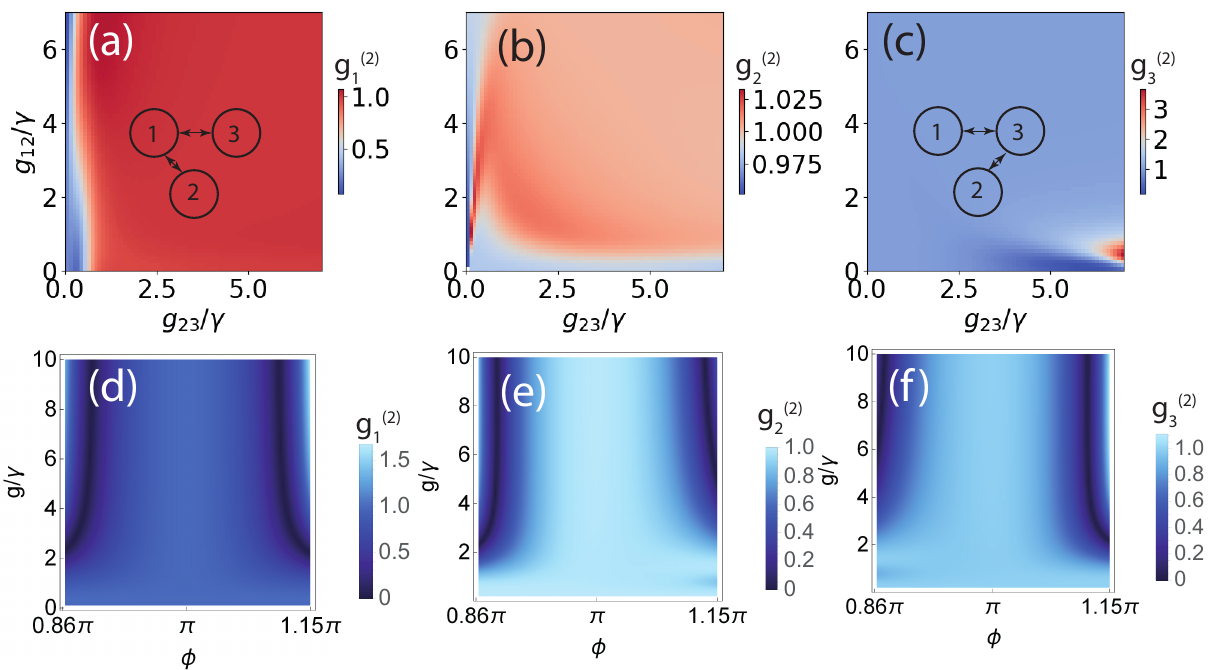}}
\caption{The $g^{(2)}$-correlation function map for the first (a), second (b) and third (c) micropillars in dependency of the coupling amplitudes between the second and the first micropillars $g_{12}$ and the second and the third micropillars $g_{23}$, other parameters are (for c-e panels) $\Delta_{1,2,3}/\gamma=0.3, F_1=0.1\gamma, F_{2,3}=0, g_{13}=g_{31}=6.2\gamma$. The $g^{(2)}$-correlation function map for the first (d), second (e) and third (f) micropillars of Hermitian symmetrical trion when $g_{23}=g_{32}=g_{12}=g_{21}=g = 6.2\gamma,\Delta_1=\Delta_2=\Delta_3=-\frac{U}{2}-\frac{g}{2}cos(\phi)$ in artificial gauge field $g_{13}=g e^{i\phi}, g_{31}=g e^{-i\phi}$. These results were obtained from the solution of the master equation (\ref{Eqm12_}) using the QuTiP package version 4.7.1, a Quantum Toolbox in Python \cite{qutip}.}
\label{fig3_2}
\end{figure}

Let’s connect another micropillar «2» to the dimer, which we denote as «1+3». With this designation, micropillar «1» is laser pumped. If we connect micropillar «2» to micropillar «1», then the blockade effect remains on it as the connection $g_{12}$ increases, see Fig.~\ref{fig3_2}a. When the second micropillar is coupled with third micropillar, then, at a certain value of the coupling parameter $g_{23}$, an antibunching effect is observed on the third micropillar, see Fig~\ref{fig3_2}c. In both cases, the statistics on the second micropillar are almost coherent $g_2^{(2)}\approx 1$, see Fig.~\ref{fig3_2}b.

If, both coupling parameters $g_{13}$ and $g_{23}$ are simultaneously increased, the antibunching effect disappears. Thus ring coupling destroys the effect of unconventional blockade. 
This is due to the fact that the cycles appears in quantum trajectories, example the antibunching disappears due to the occurrence of a cycle
$|101>\Leftrightarrow |011>\Leftrightarrow |002>$.

In the case of symmetry of the Hermitian trion (symmetry between the second and third micropillars) and when the strength of the coupling is the same the quantum statistic is almost coherent. However, if introduce the gauge filed $g \Rightarrow ge^{i\phi}$, then arise the antibunching effects, this show in figures~\ref{fig3_2}(d-f) with detunings $\Delta_1=\Delta_2=\Delta_3=\Delta=-\frac{U}{2}-\frac{g}{2}cos(\phi)$ with the breaks symmetry between the second and third micropillars see Fig.~\ref{fig3_2}(e) with left bias and Fig.~\ref{fig3_2}(f) with right bias.

\subsection{The collective blockade effect}

When we do not distinguish from which micropillar the photon came from, we can consider the whole trion or a selected part of it as a single photon source (due to the quantum  interference) \cite{Dousse2010}.

The collective blockade of the mode $a_{\sum{i}}=\sum{a_i}$ \cite{Flayac2017, Miranovitcz2010} based on quantum interference of two particle states. 
%
%
The second-order correlation function of the sum signal output from the trion is defined as:
\begin{equation} \label{Eq16b_} 
\begin{array}{l} 
{g_{1+2+3}^{\left(2\right)}(0) = \frac{\left\langle \hat{a}_{1+2+3}^{\dag 2} \hat{a}_{1+2+3}^{2} \right\rangle}{\left\langle \hat{a}_{1+2+3}^{\dag} \hat{a}_{1+2+3}\right\rangle^{2}} \approx}
\\{\approx 2\mathcal{N}\frac{|c_{200}+\sqrt{2}c_{110}+c_{020}+\sqrt{2}c_{101}+c_{002}+\sqrt{2}c_{011}|^2}{|c_{100}+c_{010}+c_{001}|^4}.} \end{array}
\end{equation}
where $\mathcal{N}=\sum_{n=0}^{M}{\sum_{i+j+k=n}{|c_{ijk}^{(n)}|^{2}}}$ is the normalization constant. 

The second order of the collective mode of a dimer «1+3» can be defined as follows (blocked radiation from second micropillar):

\begin{equation} \label{Eq16a_}
g_{1+3}^{\left(2\right)}(0) = \frac{\left\langle \hat{a}_{1+3}^{\dag 2} \hat{a}_{1+3}^{2} \right\rangle}{\left\langle \hat{a}_{1+3}^{\dag} \hat{a}_{1+3}\right\rangle^{2}} \approx 2\mathcal{N}\frac{|c_{200}+c_{002}+\sqrt{2}c_{101}|^2}{|c_{100}+c_{001}|^4}, 
\end{equation}

A polariton trion feels the collective blockade effect in the presence of a gauge field, for collective mode of whole trion «1+2+3» see Fig.~\ref{fig4} $min(g^{(2)}_{1+2+3})=0.016$. 

\subsection{non-Hermitian trion}

Finally, we will investigate the effect of non-Hermitian in the polariton trion.
The non-Hermitian coupling can increase entanglement \cite{Clerk2022}. One can generalize the Hillery-Zubairi entanglement criterion for the collective mode. The Hillery-Zubairi two mode entanglement criteria \cite{Miranovitcz2010, Hillery2006} for  one mode «2» and sum mode «1+3»:

\begin{equation} \label{Eq17EHZ_}
\begin{array}{l}
{EHZ^{(1+3,2)}=\frac{\left\langle\hat{a}_{1+3}^{\dag}\hat{a}_{1+3}\hat{a}_2^{\dag}\hat{a}_2\right\rangle}{|\left\langle\hat{a}_2^{\dag}\hat{a}_{1+3}\right\rangle|^{2}}}\\
{\approx \frac{|c_{110}+c_{011}|^2}{|c_{011}(c_{101}^*+\sqrt{2}c_{002}^*)+c_{110}(c_{101}^*+\sqrt{2}c_{200}^*)+\sqrt{2}c_{020}(c_{110}^*+c_{011}^*)|^2} \stackrel{\text{ent}}{<} 1.}\end{array}
\end{equation}


The Hillery-Zubairi entanglement criteria  for three mode (individual) of the trion is
\cite{Miranovitcz2010, Hillery2006}:

\begin{equation} \label{Eq18EHZ_}
EHZ^{(1,2,3)}=\frac{\langle \hat{a}_1^{\dag}\hat{a}_1 \hat{a}_2^{\dag}\hat{a}_2 \hat{a}_3^{\dag}\hat{a}_3 \rangle}{|\langle a_1^{\dag} a_2 a_3 \rangle|^2} \stackrel{\text{ent}}{<}1.
\end{equation}


Let us present simple considerations of the connection between quantum blockade of collective modes and quantum entanglement, with simultaneous blockade on the entire trion and on the non-Hermitian dimer, we obtain from the numerator $\eqref{Eq16b_}$ and $\eqref{Eq16a_}$ (when the blockade) that $c_{020}+\sqrt{2}(c_{110}+c_{011})\approx 0 $, then $(c_{110}+c_{011})\approx -c_{020}/\sqrt{2}$ and since $EHZ^{(1+3,2)} \sim |c_{110}+c_{011}|^2$, we have $EHZ^{(1+3,2)} \approx g^{(2)}_2$.

In region minimum $log_{10}EHZ^{(1+3,2)}$ we have small bipartite entanglement states \cite{Walter2016} $(12)|3$ for the first and second MP with $EHZ^{(1,2)}=0.9964$ and small bipartite entanglement states $2|(13)$ with $EHZ^{(1,3)}=0.9994$. This is $1-$qubit biseparable state \cite{Cirac2000}. However when quantum superposition \cite{Flayac2013} is then radiation from non-reciprocal dimer «1+3» we have $EHZ^{(1,3)}<1$. 

When blockade is simultaneously present on the trion of Fig.~\ref{fig5}e and on the non-hermitian dimer of Fig.~\ref{fig5}e, and antibunching is observed on the second micropillar of Fig.~\ref{fig5}d , then quantum entanglement between the emission of the dimer and the second micropillar is observed in the system, while three-mode quantum entanglement for the whole trion is absent, see Fig.~\ref{fig5}g.

The emission of the whole trion is almost coherent $g_{1+2+3}^{(2)}(0)\approx 1$. However, if we block the radiation from the second micropillar (associated with the nonreciprocal polariton dimer), the emission of the trion becomes more ordered than the coherent emission.

\begin{figure}
\center{\includegraphics[width=\columnwidth]{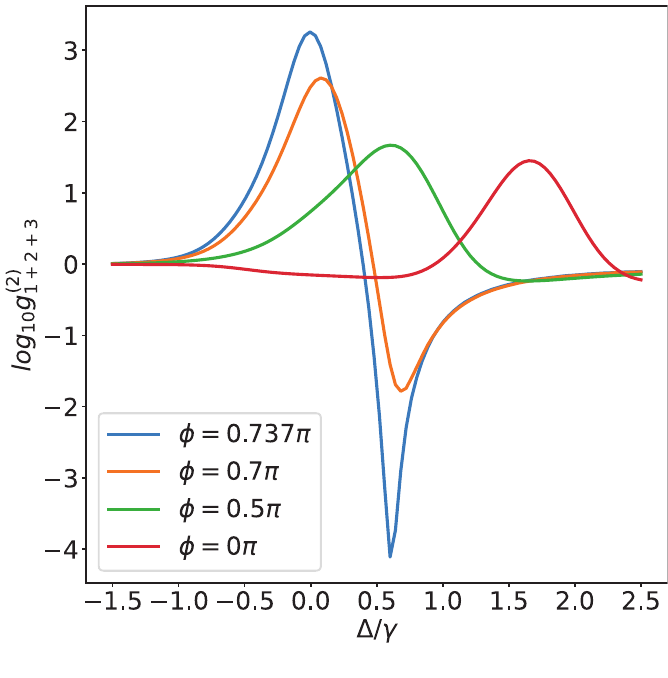}}
\caption{The second order correlation functions for collective mode $a_{1+2+3}$ $\eqref{Eq16b_}$ in dependence detuning $\Delta$ for various phases.  Parameters are: 
$g_{12}=g_{21}=6\gamma,g_{23}=g_{32}=5\gamma, |g_{13}|=\gamma$. These results were obtained with using the QuTiP package version 4.7.1, a Quantum Toolbox in Python \cite{qutip}. 
}
\label{fig4}
\end{figure}

\begin{figure}
\center{\includegraphics[width=\columnwidth]{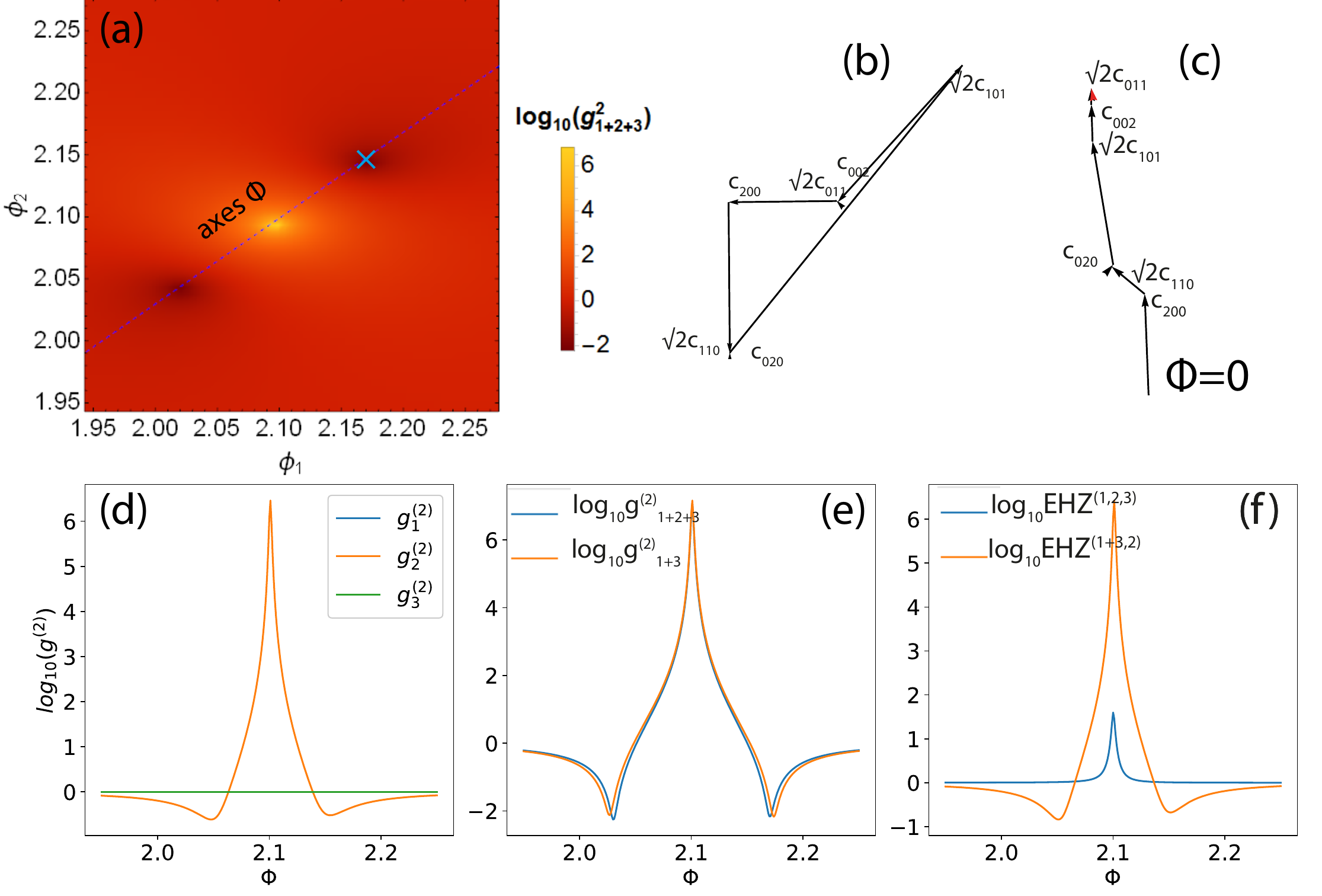}}
\caption{(a). The heat map of output mixed second order for the signal from the whole triple-micropillar structure in dependency artificial gauge fields $\phi_{1,2}$; 
(b) Graphic representation of the summation amplitude of probability for blockade case (blue cross on panel (a)) and (c) not collective blockade $\Phi=0$; (d) The second order correlation function for each MP; (e) The output mixed second order correlation function from whole structure and non-hermitian coupled MP depending on the gauge field $\Phi(\phi_1,\phi_2)$
along the dashed line in panel (a);
(f) Entanglement criterion  Hillery-Zubairi \eqref{Eq17EHZ_},\eqref{Eq18EHZ_}. Parameters are: $g_{+}=g_{-}=5\gamma,g=0.4\gamma, \Delta=-4.76\gamma$. These results were obtained from the solution of the master equation (\ref{Eqm12_}) using the QuTiP package version 4.7.1, a Quantum Toolbox in Python \cite{qutip}.
}
\label{fig5}
\end{figure}

The antibunching effect is recognized with a minimum value of $g_{1+3}^{(2)}(0)=0.058$. Figures \ref{fig4}a,b show these correlation functions as a function of the $\Delta$ detuning. Note that for individual micropillars $g_{1,2}^{(2)}(0)\approx 1$, $g_{3}^{(2)}(0)\approx 0.8$.

The collective blockade conditions are achieved if some probability amplitudes become linearly dependent, i.e., the linear combination of some complex amplitudes (of the whole set of states) is zero. The blockade condition can be represented as a condition of closure of complex probability amplitudes multiplied by the corresponding factor, see Fig.~\ref{fig5}b and no close in Fig.~\ref{fig5}c (the case of the absence of a gauge field is shown). 


%
Although the system depends on two phases, for the special case shown in the Fig.~\ref{fig5}(d-f), we can distinguish a pseudo one-dimensional phase, designated as $\Phi$, see axes $\Phi$ on the Fig.~\ref{fig5}a. In particular, if the phase is removed from the Hermitian coupling with these parameters, there will be no short circuit, as illustrated by the green arrow in Fig.~\ref{fig5}c.

Quantum blockade of the collective mode of the trion occurs in the presence of an artificial gauge field in the presence of two non-zero phases $\phi_1, \text{and} \phi_2$. Along the dashed line - one dimensional gauge field $\Phi(\phi_1, \phi_2)$ are shown on the Fig.~$\ref{fig5}$a, the effects of bunching and antibunching are manifested. 
In this gauge field in the central region in Fig.~$\ref{fig5}$(a,d), a region with a giant bunching effect (superradiation effect) appears, Fig.~$\ref{fig5}$(a,e), on both sides of which there are two regions with a strong antibunching effect with $g^{(2)}_{1+2+3}=0.006$ in the form of two spots (analogy observed in works \cite{Flayac2017} and \cite{Wang2024}).

\subsection{Invariant collective superbunching effect}


Interference of different light beams is possible if we do not distinguish from which source the photon came. However, even in this case, separating the dimer state from the whole trion is ambiguous. We can write the follow state (in two-particle basis):

\begin{equation} \label{Eq18_}
\begin{array}{l} 
{| \psi \rangle_{13,2} = \alpha_0| 0 \rangle_{13}| 0 \rangle_{2}+\alpha_1| 0 \rangle_{13}| 1 \rangle_{2}+\alpha_2| 0 \rangle_{13}| 2 \rangle_{2}+\alpha_3| 1 \rangle_{13}| 0 \rangle_{2}}\\{+\alpha_4| 1 \rangle_{13}| 1 \rangle_{2}+\alpha_5| 2 \rangle_{13}| 0 \rangle_{2},}
\end{array}
\end{equation}
%
%
here $| n \rangle_{13}$ - is state of dimer «1+3». 
For such a partition to be feasible, it is necessary that $c_{100}c_{011}=c_{001}c_{110}$, if this is not true, then it is impossible to represent the dimer state «1+3» from whole system.


\begin{figure}
\center{\includegraphics[width=\columnwidth]{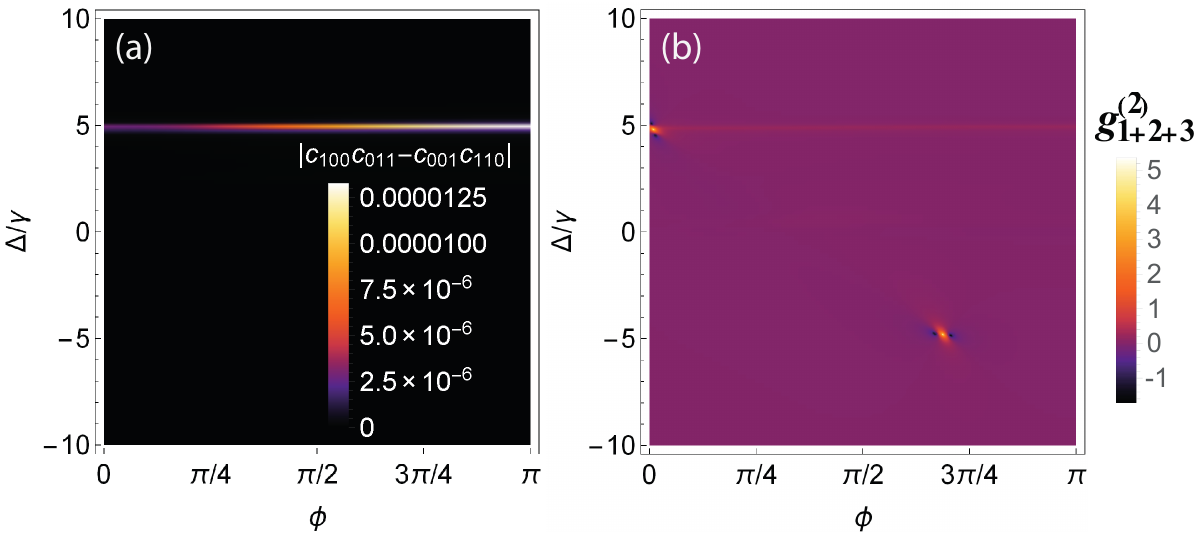}}
\caption{(a) The difference $|c_{100}c_{011}-c_{001}c_{110}|$ in dependency phase and detuning; (b) The heat map of output mixed second order for the signal from the trion micropillars in dependency artificial gauge fields $\phi_{1}=\phi_{2}=\phi$ and detuning $\Delta$. Parameters are: $g_{+}=g_{-}=5\gamma,g=0.4\gamma$.}
\label{fig6}
\end{figure}

In the region where the equality $c_{100}c_{011}=c_{001}c_{110}$ is violated see Fig.~\ref{fig6}a, the collective superbunching effect occurs of trion radiation photons is observed, see light stripe at $\Delta=5\gamma$ on the Fig.~\ref{fig6}b. This effect almost independent of the gauge phase.

\section{Conclusion}

The study shows that non-classical phenomena of the sum mode of a polariton trion occur in the presence of artificial gauge fields. For the polariton blockade effect to manifest, it is necessary for the probability amplitudes of the two-particle states to be linearly dependent.  We have shown that under the condition of antibunching of collective modes, entanglement arises between subsystems.


It is known that for indistinguishable bosons, the probability of occupying a single state increases with the population. Interparticle interactions can prevent such population and lead to the effect of quantum unconventional blockade. 
The study identified conditions under which collective blockade is observed in several micropillars connected in a closed (triangle) configuration.

We have found that collective unconventional blockade leads to quantum entanglement between subsystems of the polariton trion. It was found that when in the polariton trion the dimer state cannot be detached, the invariance of the bunching of the radiation of the system to the gauge phase is observed.

\section{Acknowledgment}
The research was carried out within the state assignment in the field of scientific activity of the Ministry of Science and Higher Education of the Russian Federation
(theme FZUN-2024-0018, state assignment of the VlSU).
The research was carried out within the state assignment in the field of scientific activity of the Ministry of Science and Higher Education of the Russian Federation
(theme FZUN-2024-0019, state assignment of the VlSU).

\section{Appendix}

\subsection{Invariant phases}

Imagine the follow Hamiltonian


\begin{equation} \label{D1_}
H=\sum_{|j-i|=1}^N g_{ij}e^{-i\phi_{ij}}\hat{a}_i^\dag\hat{a}_j+h.c.
\end{equation}
where $\phi_{ij}$ - is a phase coupling.

The annihilation operator admit a local gauge transformation that is the phase shift by $\theta$:

\begin{equation} \label{D2_}
\hat{a}_i\rightarrow \hat{a}_ie^{i\theta_i},
\end{equation}
in result is transformation of coupling amplitude as:

\begin{equation} \label{D3_}
g_{ij}\rightarrow g_{ij}e^{i(\theta_i-\theta_j)}.
\end{equation}

\begin{figure}
\center{\includegraphics[width=\columnwidth]{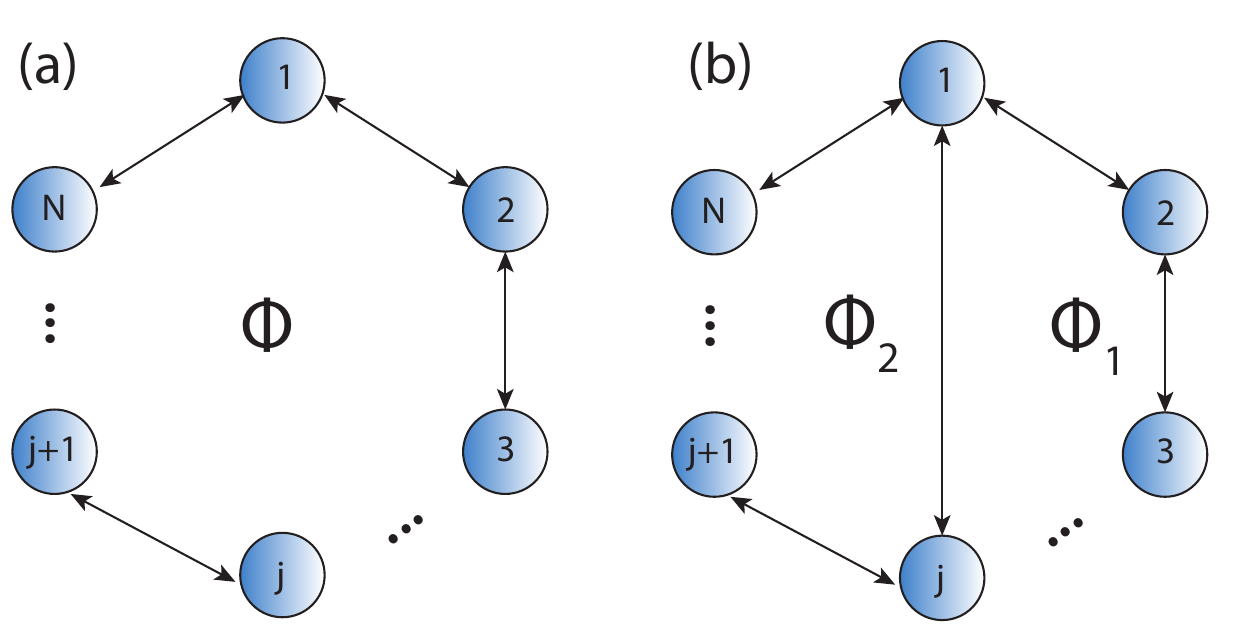}}
\caption{A Graph scheme (a) different invariant phases and (b) with jumper are connected nodes «j» and «1» phase.}
\label{figD1}
\end{figure}

For a closed loop see Fig.~\ref{figD1}a, are no gauge transformations $\eqref{D2_}$ to make all the coupling constants in the loop real numbers. However, all phases on the coupling amplitudes can be transformed by $\eqref{D3_}$, and using following procedure:

\begin{equation} \label{D4_}
\begin{array}{l} 
{\theta_2=\theta_1-\phi_1,}\\
{\theta_3=\theta_2-\phi_2,}\\
{...}\\
{\theta_{N}=\theta_{N-1}-\phi_{N-1},}
\end{array}
\end{equation}
reduced to one invariant phase $\Phi=\sum_{i=1}^N \phi_i$ (here $ \phi_i= \phi_{ii+1}$) by $g_{N1}$ coupling amplitude, this phase analogy with the non-trivial effective Aharonov-Bohm phase it's presented as ”synthetic gauge field” \cite{Clerk2022}.

The invariance of $\Phi$ can be proven as follows: suppose that there is a set of phases that can make the calibration we need. Let's write the system of linear equations and gauge transformations as the gauge equations $\theta_{j+1}-\theta_{j}+\phi_j=0$, and add all the equations, then on the left side we get $\sum_j{\phi_j}$, and on the right side is zero, we get a discrepancy.

Connected nodes «1» and «j» see Fig.~\ref{figD1}b, then we have one more equation $\theta_{j}-\theta_{1}+\phi_{N+1}=0$. It can be shown that in such a graph there is no longer one invariant phase, but two such phases.


Then if phase $\Phi$ is in this graph, then such a system is possible, in the same means as shown above. However, if we do the same on the second graph, we will get a discrepancy, since the resulting system of gauge equations is unsolvable for an arbitrary set of phases ${\phi_j}$. If the phase is on the jumper $g_{j1}$, then the system is unsolvable for the whole loop without the jumper.

\subsection{Mathematical formalism}

The corresponding wave function of the system can be
written as a ket vector in the Fock-space basis:

\begin{equation} \label{Eqt1_}
| \psi \rangle = \sum_{i+j+k<=M} c_{ijk}(t)| ijk \rangle,
\end{equation}
here $c_{ijk}$ is the amplitude of the probability of finding the system in the state $| i j k \rangle$ with $i+j+k=n$, which denotes a state with $i$ particles in the first mode, $j$ particles in the second mode and $k$ particles in the third mode. in the last equality we introduce state with $n$-particle quantum state; $M$ is the size of the truncated Hilbert space (the maximum number of particles in the system).

From the Hamiltonian \eqref{Eqm2_}, we can obtain the following system of differential equations on the state amplitudes \eqref{Eqt1_}:

\begin{equation} \label{Eqt2_}
\begin{array}{l} {\frac{id c_{ijk}}{dt}= (\Delta_{1}i+\Delta_{2}j+\Delta_{3}k}\\{+U (i(i-1)+j(j-1)+k(k-1)))c_{ijk}}\\{+F_1\sqrt{i+1}c_{(i+1)jk}+F_1\sqrt{i}c_{(i-1)jk}}\\{+F_2\sqrt{j+1}c_{(i(j+1)k}+F_2\sqrt{j}c_{i(j-1)k}}\\{+F_3\sqrt{k+1}c_{ij(k+1)}+F_3\sqrt{k}c_{ij(k-1)}}\\{+g_{12}\sqrt{i(j+1)}c_{(i-1)(j+1)k}+g_{21}\sqrt{j(i+1)}c_{(i+1)(j-1)k}}\\{+g_{13}\sqrt{i(k+1)}c_{(i-1)j(k+1)}+g_{31}\sqrt{j(i+1)}c_{(i+1)j(k-1)}}\\{+g_{23}\sqrt{j(k+1)}c_{i(j-1)(k+1)}+g_{32}\sqrt{k(j+1)}c_{i(j+1)(k-1)}.}
\end{array}
\end{equation}

Consider the processes preserving the number of particles in the system, $n$ identical particles can be placed on $m$ cells in ${n \choose n+m-1}$ ways. In the case $m=3$ we have $T_{n+1}=\frac{(n+1)(n+2)}{2}$ of different combinations. This number is the size of the $n-$partial vector $c^{(n)}$ of probability amplitudes:

\begin{equation} \label{Eqt3_}
\begin{array}{l} {c^{(n)}=(c_{0,n,0},c_{1,n-1,0},c_{0,n-1,1},c_{2,n-2,0},c_{1,n-2,1},c_{0,n-2,2},}\\{...,c_{(m-(i-t_m(i)),(n-m),(i-t_m(i))},...,c_{0,0,n})^{T}}
\end{array}
\end{equation}

The vector \eqref{Eqt3_} consists of the state amplitudes of all possible unique combinations of finding a system with $n$ particles distributed over the system. These combinations include different partitions of the $n$ particles into different numbers of particles in each of the $n_{1,2,3}$ cells, giving a total of $n$ and different permutations of the unique sum over the cells. For the latter case, there are three possibilities: 1) when $n_1=n_2=n_3=n'$, in which case there can be only one combination when the number of particles is a multiple of 3 $(n',n',n')$, where $n'=n/3$ with the corresponding probability amplitude $c_{n',n',n'}$; 2) when $n_i=n_j=n''\neq n_k$, in which case we have $3[\frac{n+1}{2}]$ of such variants with group $(c_{n'',n'',n_3},c_{n'',n_2,n''},c_{n_1,n'',n'',n''})^{T}$ and the third case, when the numbers $n_i$ are completely distinct, then the permutation group has six elements $(c_{n_1,n_2,n_3},c_{n_2,n_3,n_1},c_{n_3,n_1,n_2},c_{n_2,n_1,n_3},c_{n_1,n_3,n_2},c_{n_3,n_2,n_1})^{T}$.

\begin{figure}
\includegraphics[width=0.49\textwidth]{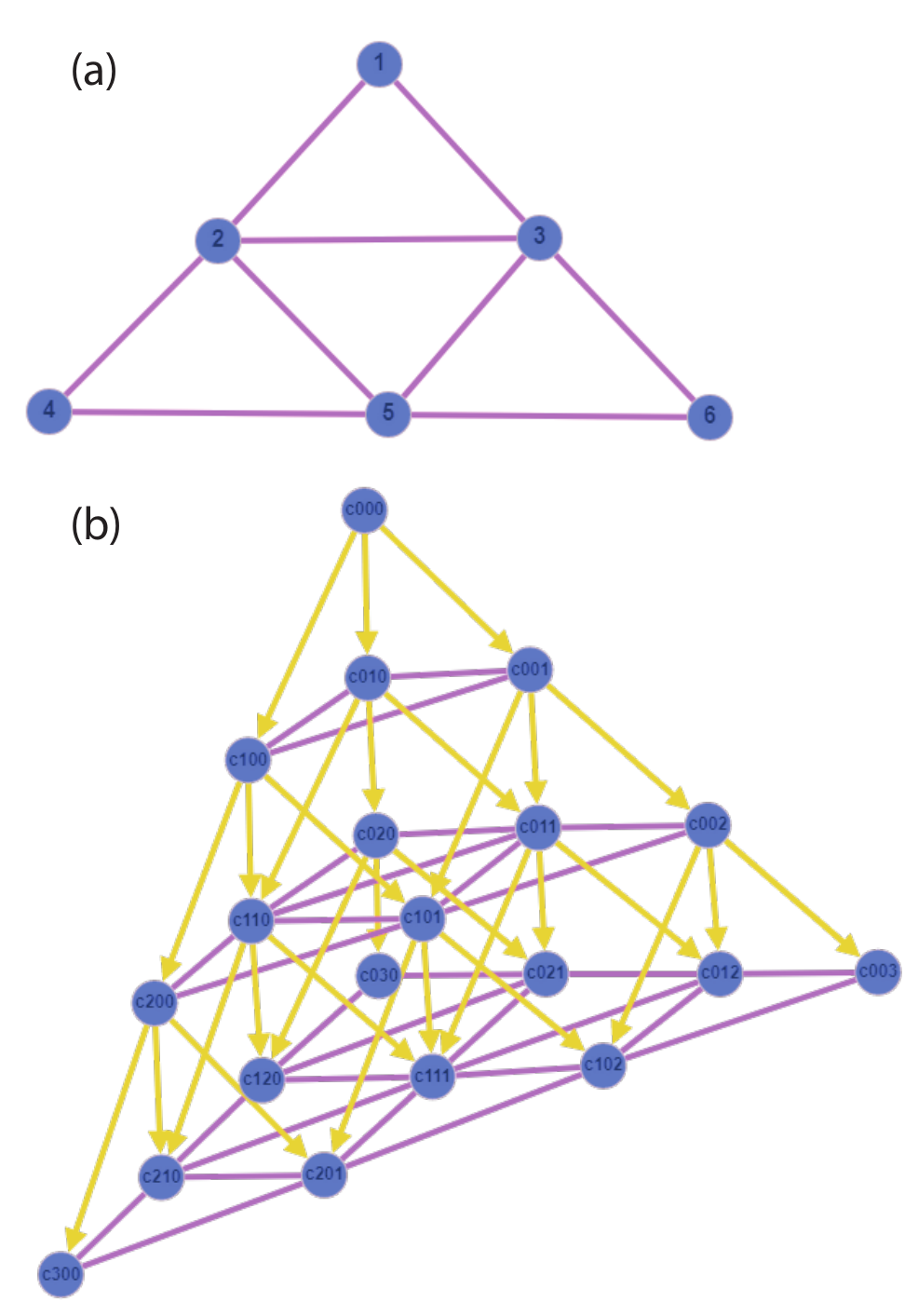}
\caption{(a) A sketch of the Fock two-particle tensor state space of a trion. (b) Part of the Fock space of states of the trion. Nodes correspond to different states, graphs with purple edges correspond to $n-$particle tensor states, yellow arrows show transitions between different tensor states.}
\label{fig7}
\end{figure}

The Fock space for an $n-$-particle state is a triangular lattice graph $G(V,E)=TG_{n}$, see Fig.~\ref{fig7}a, by connecting each tensor state to each other (by pumping, yellow arrows in Fig.~\ref{fig7}(b) we get a complete representation of the trion state, see Fig.~\ref{fig7}b. The set of vertices for a triangular lattice graph is determined from the rule: $V=\left\{(i,j,k): i+j+k=n, i,j,k\geq0\right\}$, two vertices are connected only if $E=\left\{(e_1,e_2):|i_{e_1}-i_{e_2}|+|j_{e_1}-j_{e_2}-j_{e_2}|+|k_{e_1}-k_{e_2}|=n \right\}$.

The bunching of elements in \eqref{Eqt3_} was carried out as follows. Let us imagine a triangular lattice with elements $T_3$, see Fig.~\ref{fig7}a, this is a representation of a tensor state with $n=2$ particles. At the top we will have the state $c_{0,2,0}$, going down to the lower level, we decrease the number of particles in the central cell by one, giving it either to the left neighboring cell if we move to the left, or to the right cell if we move to the right. Let us number all elements of the lattice from top to bottom and from left to right. Let us divide the entire set of ordinal numbers of the triangular lattice into levels (subsets), as shown in Fig.~\ref{fig7}b. The following is true: if a number $a$ is greater than $T_{m-1}$ and less than (or equal to) $T_{m}$, then this number is at the $m$-th level. The number $i$ corresponds to the state $c_{(m-k),(n-m),k}$. Here $m=1_m(i)$, where $1_m(i)$ is the indicator function. The index $k$ is defined as follows: $k=i-T_{m-1}$.

Following \cite{Wang2024} and neglecting quantum jumps, we can write the equation governing for the time dependency n-particles state $| \psi^{(n)}(t) \rangle$ for the non-Hermitian Hamiltonian:

\begin{equation} \label{Eqt4_}
i\frac{d}{dt}| \psi^{(n)}(t) \rangle  \approx H_{0} | \psi^{(n)}(t) \rangle + H_{+} | \psi^{(n-1)}(t) \rangle,
\end{equation}
here we introduce the Hamiltonian that does not change the number of particles $H_{0}=\hat{a}^{\dag{}}\mathbf{G_{eff}}\hat{a}+\sum_i{U_{i}\hat{a}^{\dag{}}_{i}\hat{a}^{\dag{}}_{i}\hat{a}_{i}\hat{a}_{i}}$ and increases the number of particles by one $H_{+}=\sum_i{F_{i}\hat{a}_{i}^{\dag{}}}$ and neglecting the annihilation part of the pump Hamiltonian for weak pump limit due to sub-leading ordered $c^{(n+1)} \ll c^{(n)}$ \cite{Wang2024}. 

From \eqref{Eqt4_} We can obtain the follow differential equations by amplitudes:

\begin{equation} \label{Eqt5_}
\frac{dc^{(n)}}{dt}=G^{(n)}c^{(n)}+f^{(n)}c^{(n-1)},
\end{equation}
where we introduce the $n$-state amplitude vector $c^{(n)}=(c_{n,0,0},c_{n-1,1,0},c_{n-1,0,1},...,c_{n-i,i,0},c_{n-i,i-1,1},...,c_{n-i,i-j,j},\\
...,c_{n-i,0,i},...,c_{0,0,n})^{T}$ (which are contain $T_{n+1}$ elements, where  $T_{n} = \frac{(1 + n) n}{2}$ triangular numbers). Here we introduce $n-$particle quadratic form matrix:

\begin{equation} \label{Eqt6_}
\begin{array}{l}{G^{(n)}_{i,i'}=\sum_{j=1,2,3}{(\Delta_j t(i)_j+U t(i)_j(t(i)_j-1)) \delta_{i,i'}}}\\{+[i>1 \wedge i \neq T_m, \forall m, \wedge i \neq T_{m}+1 \wedge i<T_{n-1}]}\\{(g_{12}\sqrt{t_1(t_2+1)}\delta_{i',T_{n-t_2-1}+t_3+1}}\\{+g_{21}\sqrt{t_2(t_1+1)}\delta_{i',T_{n-t_2+1}+t_3+1}}\\{+g_{13}\sqrt{t_1(t_3+1)}\delta_{i',T_{n-t_2}+t_3+2}}\\{+g_{31}\sqrt{t_3(t_1+1)}\delta_{i',T_{n-t_2}+t_3}}\\{+g_{23}\sqrt{t_2(t_3+1)}\delta_{i',T_{n-t_2+1}+t_3+2}}\\{+g_{32}\sqrt{t_3(t_2+1)}\delta_{i',T_{n-t_2-1}+t_3})}\\{+\delta_{i',2}g_{21}\sqrt{t_2}}\\{+\delta_{i',3}g_{23}\sqrt{t_3}}\\{+\sum_{m=2}^{n-1}{\delta_{i,T_m}g_{21}\delta_{i',T_m+m}\sqrt{n-m+1}}}\\{+g_{23}\delta_{i',T_m+m+1}\sqrt{m(n-m+1)}}\\{+g_{31}\delta_{i',T_{m}+m-1}\sqrt{m-1}}\\{+g_{32}\delta_{i',T_{m-1}+m-1}\sqrt{(m-2)(n-m+2)}}\\{+\sum_{m=1}^{n-2} \delta_{i',T_m+2}g_{13}\sqrt{m}}\\{+\delta_{i',T_{m+1}+2}g_{23}\sqrt{n-m}}\\{+\delta_{i',T_{m-1}+1}g_{12}\sqrt{m(n-m+1)}}\\{+\delta_{i',T_{m+1}+1}g_{21}\sqrt{(n-m)(m+1)}}\\{+\delta_{i',T_{n-1}+1}g_{12}\sqrt{n}}\\{+\delta_{i',T_n+2}g_{13}\sqrt{n}}\\{+(\delta_{i',i-1}g_{31}\sqrt{(i-T_n)(n-i+T_n+1)}[T_n+1<i'<T_n+1]}\\{+\delta_{i',i+1}g_{13}\sqrt{(n-i+T_n)(1+i-T_n)}}\\{+\delta_{i',i+T_{n-1}-T_n-1}g_{32}\sqrt{i-T_n}}\\{+\delta_{i',T_{n-1}-T_n+i+1}g_{12}\sqrt{T_n+n-i})},\end{array}
\end{equation}
where $[.]$ denotes an Iverson bracket, it returns 1 if the expression in the brackets is true and 0 if it is false. The index $m$ depends on the row index of the matrix, as follows:
$m(i)=\sum_{j=1}^{n+1}j\theta(i-T_{j-1})\theta(T_j-i)$. Here $\theta(x)$ - a Heaviside step function.

The first line in \eqref{Eqt6_} describes the diagonal elements, they do not change the distribution of particles in the system. The second element defines transitions from states in the core of the Fock space. The third element corresponds to the transition from vertex $c_{0,n,0}$. The fourth and fifth elements are from states a of the right and left sides of the triangle. The sixth and seventh from the outermost states and the eighth from the bottom side of the triangle. Although such a notation does not clarify the features of matrix $G^{(n)}$, but it gives a way to write similar matrices of general form, with the help of the derived formula, we can write matrix $G^{(n)}$ for any $n$ study its structure by computer methods.

In \eqref{Eqt5_} we also introduced a matrix $f^{(n)}$ translating from an $n-$partial tensor state to a $(n+1)-$particle tensor state of size $T_{n}$ by $T_{n+1}$.
We also introduced a triangular lattice coordinate function that takes the cell number in the triangular lattice as input, and returns three amplitude indices $c_{i,j,k}$:

\begin{equation} \label{Eqt7_}
t(i)=(n-i+T_{m-1}-1,m,i-T_{m-1}+1),
\end{equation}

For example, for two-particle states, these matrices will look as follows:
\begin{widetext}
\begin{equation} \label{Eqt8_}
G^{(2)}= \begin{bmatrix} 2U + 2 \Delta_2 & \sqrt{2} g_{21} & \sqrt{2} g_{23} & 0 & 0 & 0 \\ \sqrt{2} g_{12} & \Delta_1 + \Delta_2 & g_{13} & \sqrt{2} g_{21} & g_{23} & 0 \\ \sqrt{2} g_{32} & g_{31} & \Delta_2 + \Delta_3 & 0 & g_{21} & \sqrt{2}g_{23} \\ 0 & \sqrt{2} g_{12} & 0 & 2U + 2 \Delta_1 & \sqrt{2} g_{13} & 0 \\ 0 & g_{32} & g_{12} & \sqrt{2} g_{31} & \Delta_1 + \Delta_3 & \sqrt{2}g_13 \\ 0 & 0 & \sqrt{2} g_{32} & 0 & \sqrt{2} g_{31} & 2U + 2 \Delta_3 \end{bmatrix}
\end{equation}
\end{widetext}

and
%
\begin{equation} \label{Eqt9_}
f^{(2)}= \begin{bmatrix} \sqrt{2} F_2 & 0 & 0 \\ F_1 & F_2 & 0 \\ F_3 & 0 & F_2 \\ 0 & \sqrt{2} F_1 & 0 \\ 0 & F_3 & F_1 \\ 0 & 0 &\sqrt{2} F_3 \end{bmatrix}
\end{equation}

We can find the following stationary solutions of the equations \eqref{Eqt4_} to the probability amplitudes $c^{(n)}$ by recurrently solving them for each $n=1,2,3...$:

\begin{equation} \label{Eqt10_}
\begin{array}{l}{c^{(1)}=-G_{eff}^{-1}f^{(1)},}\\
{c^{(2)}=-G^{{(2)}^{-1}}f^{(2)}c^{(1)}=G^{{(2)}^{-1}} f^{(2)} G_{eff}^{-1}f^{(1)}c^{(1)},}\\
{...,}\\
{c^{(n)}=- G^{{(n)}^{-1}}f^{(n)}c^{(n-1)}=(-1)^n\prod\limits_{i=n}^{1} G^{{(i)}^{-1}}f^{(i)}.}
\end{array}
\end{equation}

\end{document}